\documentclass[usenatbib]{mn2e}

\newcommand{\Pradio} {$P_{\rm 1.4 \, GHz}\,$}
\newcommand{\PMcorrelation} {$P_{\rm 1.4 \, GHz} - M_{500} \;$}
\newcommand{\Smt} {S^{\rm mock,tot}_{R_{\rm H}}}
\newcommand{\Smm} {S^{\rm mock,meas}_{R_{\rm H}}}
\newcommand{\Smme} {S^{\rm mock,meas}_{2\sigma}}
\newcommand{\Dmm} {D^{\rm mock,meas}_{2\sigma}}

\usepackage{graphicx}

\usepackage{subfig}
\title[On the absence of radio halos in clusters with double relics]{On the absence of radio halos in clusters with double relics}
\author[A. Bonafede  et al.] {A. Bonafede,$^{1,2}$\thanks{E-mail:email@address} 
R Cassano$^{2}$, M. Br{\"u}ggen$^{1}$, G. A.  Ogrean$^{3}$, C.J. Riseley$^4$, V. Cuciti$^{2,5}$, \\
\newauthor F. de Gasperin$^{6}$, N. Golovich$^{7}$,  R. Kale$^{8}$, T. Venturi$^{2}$,  R. J van Weeren$^{9}$, D. R.  Wik$^{10,11,12}$\\
\newauthor D. Wittman$^{5}$\\
$^1$ Hamburger Sternwarte, Universit\"at Hamburg, Gojenbergsweg 112, 21029, Hamburg, Germany. \\
$^2$ INAF IRA, via Gobetti 101, 40129 Bologna, Italy.\\
$^3$ Department of Physics, Stanford University, 382 Via Pueblo Mall, Stanford, CA 94305-4060, USA \\
$^4$ CSIRO Astronomy \& Space Science, 26 Dick Perry Avenue, Kensington, WA, 6151, Australia\\
$^5$  Dipartimento di Fisica e Astronomia, Universitˆ di Bologna, via P. Gobetti 93/2, 40129, Bologna, Italy\\
$^6$ Leiden Observatory, Leiden University, P.O. Box 9513, 2300 RA Leiden, the Netherlands.\\
$^7$ University of California, One Shields Avenue, Davis, CA 95616, USA\\
$^8$ National Centre for Radio Astrophysics, Tata Institute of Fundamental Research, Post Bag 3, Ganeshkhind, PUNE 411 007, INDIA\\
$^9 $Harvard-Smithsonian Center for Astrophysics, 60 Garden Street, Cambridge, MA 02138, USA. \\
$^{10}$ Department of Physics and Astronomy, University of Utah, 115 South 1400 East, Salt Lake City, Utah 84112, USA.\\
$^{11}$ Goddard Space Flight Center,  Greenbelt, MD 20771, USA.\\
$^{12}$ The Johns Hopkins University, Homewood Campus, Baltimore, MD 21218, USA.
}

\begin{document}

\date{Accepted  Received...}

\pagerange{\pageref{firstpage}--\pageref{lastpage}} \pubyear{2002}

\maketitle

\begin{abstract}
Pairs of radio relics are believed to form during cluster mergers, and are best observed when the merger occurs in the plane of the sky.
Mergers can also produce radio halos, through complex processes likely linked to turbulent re-acceleration of cosmic-ray electrons. However, only some clusters
with double relics also show a radio halo. Here, we present a novel method to derive upper limits on the radio halo emission,
and analyse archival X-ray Chandra data, as well as galaxy velocity dispersions and lensing data, in order to understand the key parameter 
that switches on radio halo emission. We place upper limits on the halo power below the \PMcorrelation correlation for some clusters, 
confirming that clusters with double relics have different radio properties. Computing X-ray morphological indicators, we find that
clusters with double relics are associated with the most disturbed clusters. We also investigate the role of different mass-ratios and time-since-merger. Data do not indicate that the merger mass ratio has an impact on the presence or absence of radio halos (the null hypothesis that the clusters belong to the same group cannot be rejected). However, the data suggests that the absence of radio halos could be associated with early and late mergers, but the sample is too small to perform a statistical test.
Our study is limited by the small number  of clusters with double relics. Future surveys with LOFAR, ASKAP, MeerKat and SKA will provide larger samples to better address this issue.

\end{abstract}

\par
\section{Introduction}
Cosmological simulations show that mergers between massive clusters produce shock waves that travel from the cluster cores out to  the cluster
periphery \citep{Bruggen11,Vazza12,sk13}. Shock waves heat the thermal gas in the Intracluster Medium (ICM) and produce temperature and X-ray surface brightness discontinuities 
that have been observed in some clusters \citep[e.g.][]{OgreanCiza}. At the same time, shock waves could amplify the ICM magnetic field and (re)accelerate relativistic electrons, producing synchrotron emission in the radio band. 
Radio relics are arc-shaped radio sources found at the periphery of galaxy clusters that could result from shock-(re)acceleration of fossil electrons from AGN  \citep[e.g.][]{Bonafede14a,vanWeeren17_Nature}.
Although the (re)acceleration process is not understood yet \citep{KangRyu11,KangRyu12,Pinzke13,VazzaBruggen13,Wittor17}, a connection between shock waves and radio relics is established. According to numerical and cosmological simulations, two symmetric relics can be visible when the merger axis is in the plane of the sky \citep[e.g.][]{vanWeeren11_sim,Bruggen11}. \par

Mergers between massive galaxy clusters can also produce another type of Mpc-size radio source, named radio halos \citep[see reviews by][and ref. therein]{feretti12,BJ14}. Radio halos are extended radio sources, that appear to fill the central regions of clusters with linear sizes that range from $\sim$ 500 kpc up to more than 1 Mpc. 
Despite significant progress in recent years, the origin of radio halos is not yet understood.
They could be produced by  the re-acceleration of a seed population of relativistic particles due to turbulent motions that develop in the ICM during mergers \citep{Brunetti01,Fujita03}. Although data seem to fit the theoretical picture, many open issues remain, such as the origin of the seed electrons, the (re)acceleration mechanism, and the role of the magnetic field.
Since their first discovery, it has been noticed that radio halos always occur in massive and merging clusters \citep[e.g.][]{Buote01,Venturi08}, and it has been suggested that the fraction of clusters with halos increases with the cluster mass \citep[e.g.][]{Giovannini99,Venturi08}. Using data from Planck catalogues, \citet{Cuciti15} demonstrated that this conclusion can be statistically confirmed for clusters with masses $M_{500} > 6 \times 10^{14} M_{\odot}$.  \par
A total of 17 clusters with double relics have been discovered so far (see Tab. \ref{tab:sample} and references in the caption). Only in a fraction of these clusters, though, has a radio halo been detected.  If both radio halos and relics are produced during mergers, clusters with relics and without halo permit to investigate the merger parameters and the cluster properties that determine whether or not a radio halo develops during cluster mergers. \par
Here, we focus on clusters that host double relics, as they define a sample of objects where the merger is seen close to the plane of the sky.  This minimises projection effects and permits estimates of the time since merger.\par
Radio observations of clusters with double relics have different sensitivities and have been performed with different instruments. In addition, the radio power of halos correlates with the cluster mass \citep[e.g.][]{Cassano13}, thus biasing against less massive systems. As a first step, one needs to establish whether present observations would have been able to detect a radio halo. Once a true upper limit on the radio halo emission is established, we analyse the cluster and merging properties using X-ray  data and literature information. Our aim is to search for a parameter or a combination of parameters that determines the occurrence of radio halos during mergers. \par
The paper is structured as follows: In Sec. \ref{sec:radio} we present a new method to derive upper limits on the cluster radio emission. The dynamical status of the clusters
is analysed in Sec. \ref{sec:X} using X-ray observations, and in Sec. \ref{sec:Discussion} we discuss different parameters that could cause the radio emission. Finally, we present our conclusions in Sec. \ref{sec:Conclusions}. Throughout this paper, we assume a $\Lambda$CDM cosmological model with $H_0=$69.6 km/s/Mpc , $\Omega_M=$0.286, $\Omega_{\Lambda}=$0.714 \citep{2014ApJ...794..135B}.
\par

\section{Radio halos and upper limits}
\label{sec:radio}

\subsection{The cluster sample}
\label{sec:sample}
From the literature, we have collected  the data of all clusters known to date that host  symmetric double radio relics. The main properties of these clusters are listed in Table \ref{tab:sample}.
In total, there are six clusters that are known to host double relics and no halo (top of Table  \ref{tab:sample}), seven clusters  that host, both, double relics and a radio halo (mid-top  of Table \ref{tab:sample}), 2 clusters that host one relic and one candidate counter-relic (mid-bottom of Table  \ref{tab:sample}), and two controversial cases (bottom of Table  \ref{tab:sample}). These clusters are Abell 3667 and 
ZwCl2341.1+0000. Abell 3667 hosts diffuse emission classified as ``radio bridge connecting  the NW relic of the galaxy cluster Abell 3667 to its central regions" \citep{Carretti13}, and more recently re-classified as a candidate mini halo \citep{Riseley15}. For the cluster ZwCl2341.1+0000 different interpretations have been given (see e.g \citealt{Bagchi02,vanWeeren09_ZwCl2341,Giovannini10_ZwCl2341}) and the presence of a radio halo is uncertain.
For these reasons, ZwCl2341 and A3667 have been excluded from our analysis. Hence, the final sample consists of 15 clusters. We note that the clusters span a wide range of redshifts (from $z=0.0456$ to $z=0.87$) and masses (from $\rm{M_{500}=1.2 \times 10^{14} M_{\odot}}$ to $\rm{M_{500}=14 \times 10^{14} M_{\odot}}$). \par
In this work, we use the calibrated datasets of clusters with double relics and no detected radio halo. Most of the observations were performed at 1.4 GHz. The only exceptions are  A3376 and MACSJ0025 that have been observed at 325 MHz. Details about the observations and data reduction can be found in the papers listed in Table \ref{tab:sample}.

\begin{table*}

 \centering
 \caption{Clusters with double relics.}
 \begin{tabular}{l|c|c|c c c c}
  \hline
Cluster name 			& Redshift &  Cluster mass &  Mass ref. &  Radio halo & Radio ref   & Chandra   \\
					&				&  $10^{14} \, \rm{M_{\odot}}$                &                 &                \\
&&&&&\\
PSZ1 G096.89+24.1		& 0.3			&	4.7$\pm $0.3& $^1$             & NO &WSRT 1.4 GHz $^6$  & \\
   ZwCl0008.8+5215 		& 0.104		&	3.4$\pm $0.3&	 $^1$           & NO &	WSRT 1.4 GHz$^7$    & Y\\
Abell 2345			& 0.1765		&	5.9$\pm$ 0.4&	 $^1$           & NO & VLA 1.4 GHz$^8$   & Y\\
   Abell 1240				& 0.1948		&	3.7$\pm$0.4&	 $^1$   & NO & VLA 1.4 GHz$^8$   & Y\\
  Abell 3376			& 0.0456		&	2.4$\pm$0.2&	 $^1$           & NO & GMRT 325 MHz$^9$  & Y \\
MACSJ0025.4-1222		& 0.5857		&	8.4$\pm$3.2& 	$^2$           &NO	& GMRT 325 MHz$^2$  & Y \\
&&&&&\\

8C 0212+703			& 0.0655		& 1.1$\pm$0.2*		& $^3$ & RH& $^3$  & Y \\
  El Gordo		& 0.8700		& 10.8 $\pm$0.5&     $^1$   &  RH &  $^{10}$ & Y \\
  PLCKG287.0 +32.9		& 0.39	& 14.7$\pm$0.4&      $^1$ &  RH & $^{11}$  &\\
RXCJ1314.4-2515		&0.24740		& 6.7 $\pm$0.5 &	 $^1$ &  RH&$^{12}$  & \\
MACSJ1752.0+4440		& 0.366		& 6.7$^{+0.4}_{-0.5}  $  &	 $^1$&  RH & $^{13}$  &\\
PSZ1G108.18-11.53		& 0.336		& 7.7 $\pm$0.6 &	 $^1$&  RH&$^{14}$  & \\
  CIZAJ2242.8+5301		& 0.1921		& 17.0$\pm$3.6&       $^4 $& RH &  $^{15}$ & Y\\

&&&&&\\
  Abell 3365			& 0.0926		&	1.48$\pm$0.04	& $^5$ &  DR[c] - NO& $ ^{16}$  & \\
  MACSJ1149.5+2223		& 0.5444		&	10.4$\pm$0.5  &$^1$ & DR[c] - RH &   $^{17}$& Y \\
&&&&&\\

Abell 3667			& 0.0556		&   7.04$\pm$ 0.05 &	 $^1$  &  Mini halo[c]& $^{18}$  &\\
ZwCl2341.1+0000		&0.27	& 5.2 $\pm$ 0.4  &  $^1$ & Controversial&  $^{19}$  &\\	
\hline

\multicolumn{7}{l}{\scriptsize   Col. 1: Cluster name, Col. 2: cluster redshift, Col. 3: Cluster mass, $M_{500}$ is given for all cluster except for 8C 0212+703	 and A3365. }\\
\multicolumn{7}{l}{\scriptsize   Col. 4: Reference for the cluster mass in Col. 3, Col5: Radio emission from the cluster: RH for clusters with a radio halo,}\\
\multicolumn{7}{l}{\scriptsize NO for clusters with no radio halo. Col 6: reference for the radio data in Col. 5, Col 7: Archival Chandra data.}\\
\multicolumn{7}{l}{\scriptsize References: $^1$ Planck Union Catalog \citet{PUC}, $^2$ \citet{Riseley17},  $^3$ the X-ray luminosity is taken from \citet{Brown11},   }\\
\multicolumn{7}{l}{\scriptsize considering a mean between the total X-ray luminosity and the one in the halo region. The mass is derived using the correlation by \citet{Pratt09}. }\\
 \multicolumn{7}{l}{\scriptsize $^4$ Hoang et al, submitted,  $^5$ ROSAT using \citet{Pratt09} relation. $^6$ \citet{deGasperin14}, $^7$ \citet{vanWeeren_ZwCl0008}, $^8$ \citet{Bonafede09a}, }\\
\multicolumn{7}{l}{\scriptsize   $^9$ \citet{Kale12}, $^{10}$\citet{Lindner14}, $^{11}$ \citet{Bonafede14a}, $^{12}$ \citet{Feretti05}, $^{13}$ \citet{vanWeeren12}, $^{14}$ \citet{deGasperin15},}\\
\multicolumn{7}{l}{\scriptsize  $^{15}$ \citet{vanWeeren10} $^{16}$ \citet{vanWeeren09}, $^{17}$ \citet{Bonafede12}, $^{18}$ 
\citet{Riseley15},  $^{19}$ \citet{vanWeeren09_ZwCl2341} and \citet{Giovannini10_ZwCl2341}}

\end{tabular}
\label{tab:sample}
\end{table*}


\begin{figure*}
\subfloat[]{\includegraphics[width = 8cm]{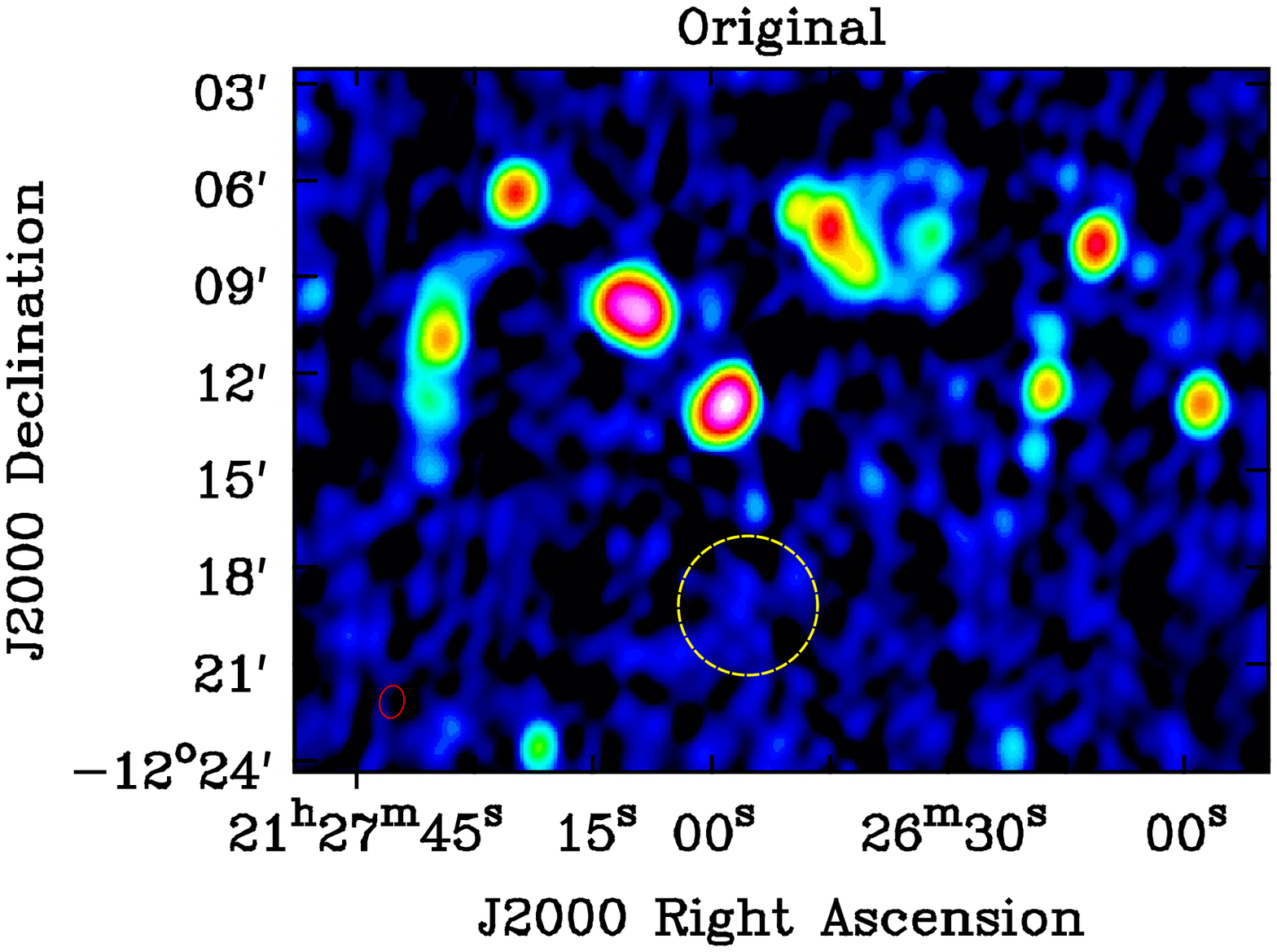}}
\subfloat[]{\includegraphics[width = 8cm]{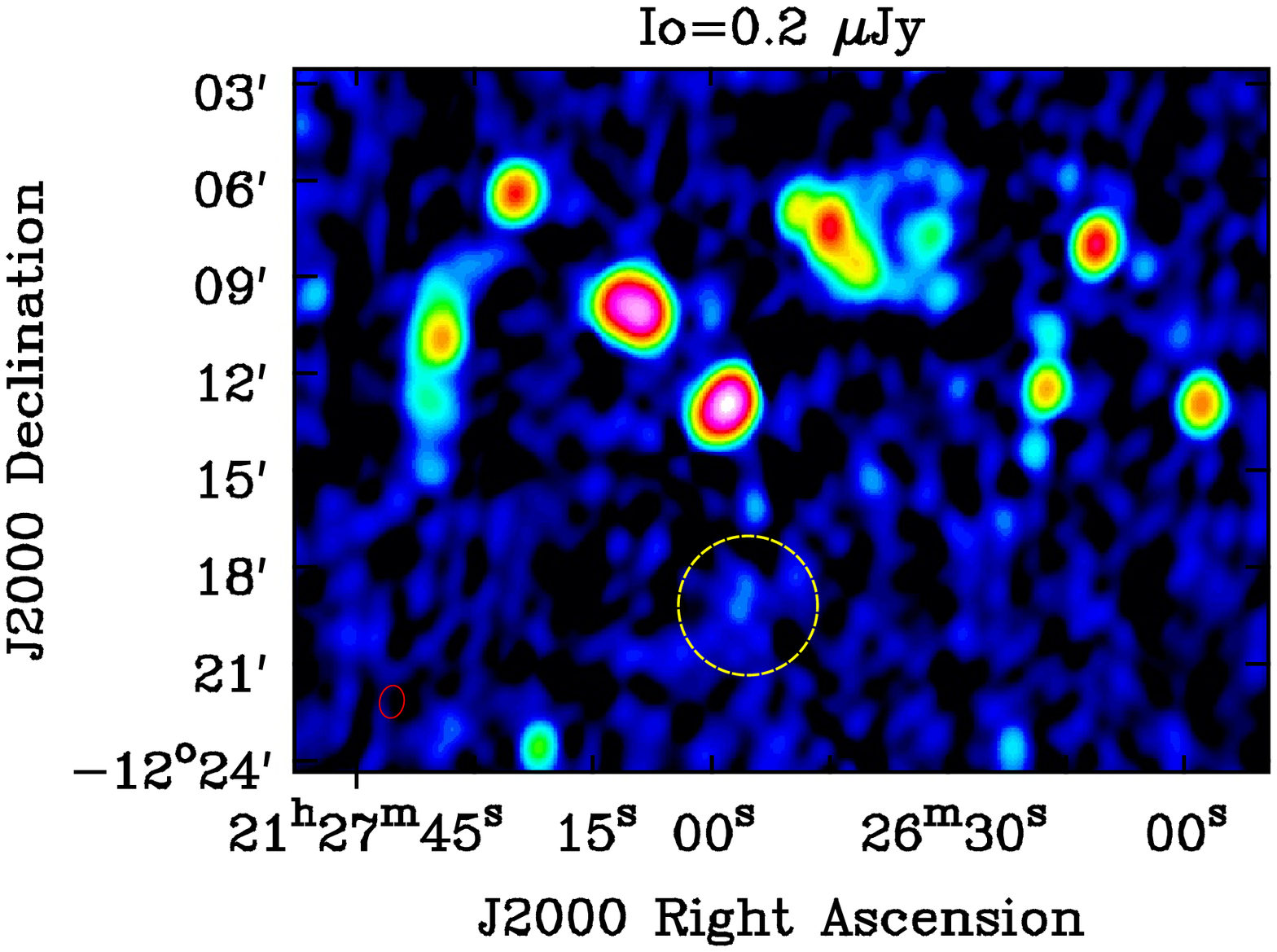}}\\
\subfloat[]{\includegraphics[width = 8cm]{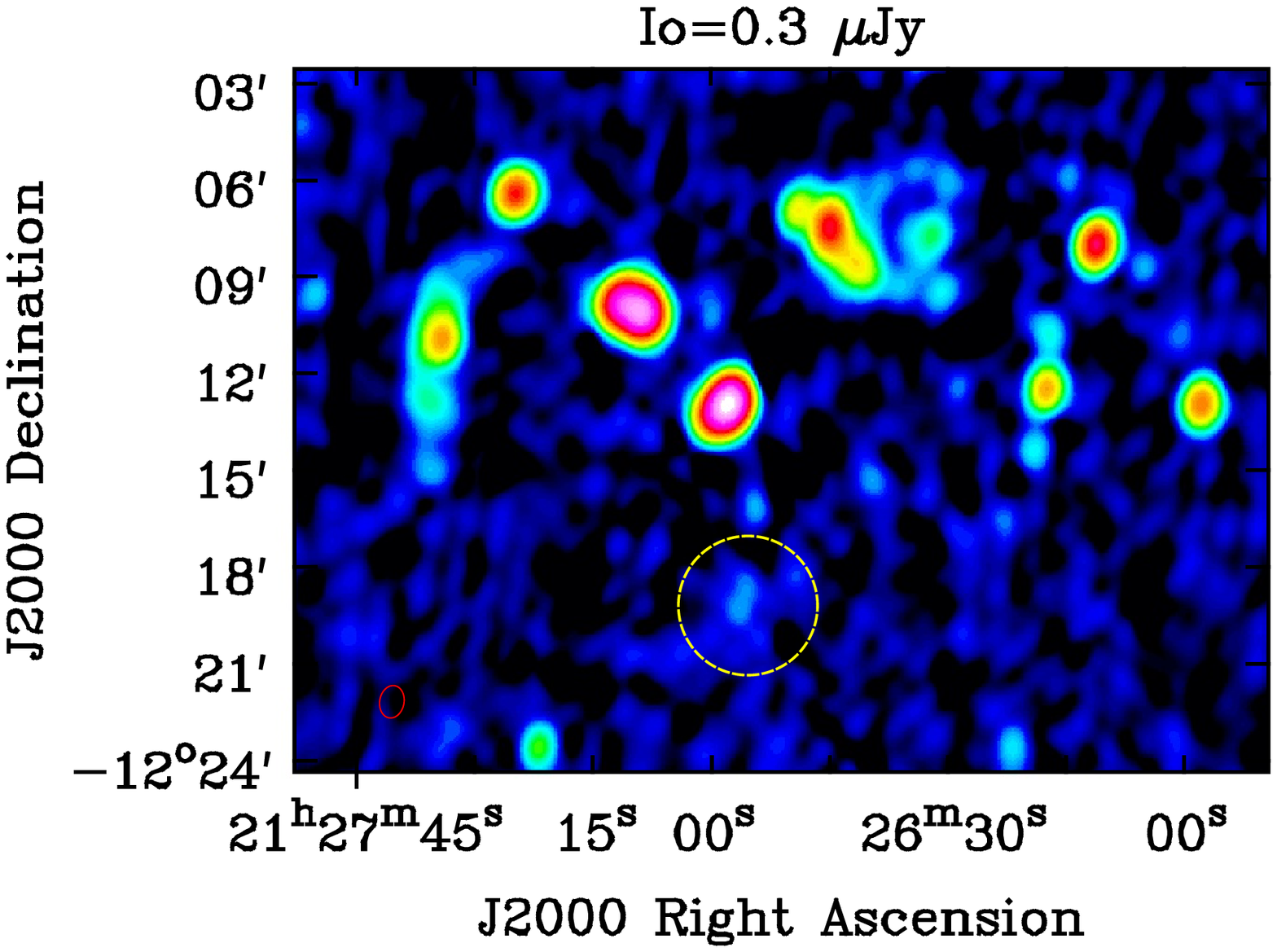}}
\subfloat[]{\includegraphics[width = 8cm]{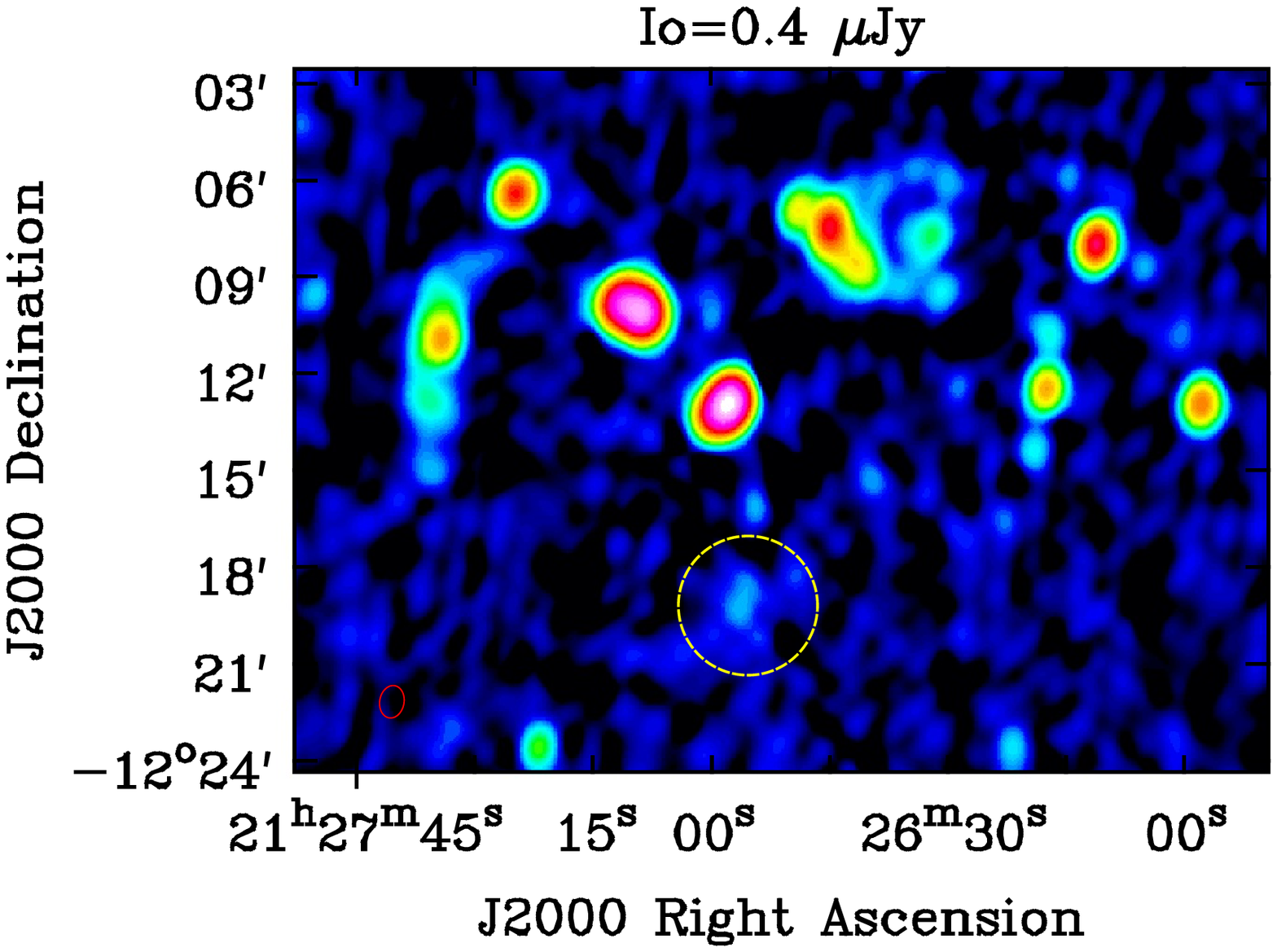}}
\caption{Mock halos in the cluster A2345. Images of the cluster (a): original dataset, (b): mock halo with $I_0=0.2 \, \mu$Jy/arcsec$^2$ (c) mock halo with $I_0=0.3 \mu$ Jy/arcsec$^2$ (d) mock halo with $I_0=0.4 \mu$ Jy/arcsec$^2$ . The yellow circle is centred on the halo and has a radius $r= R_{\rm H}$. The images have been obtained with the same parameters, and displayed with the same color scheme. The mock halo that fulfils the criteria for detection is the one in panel (c)}
\label{fig:A2345}
\end{figure*}

\subsection{A new method to derive upper limits}
\label{sec:UL}

Radio halos have a low surface brightness at frequencies higher than few hundred MHz, and the observations that we analyse in this work have different sensitivities. 
Hence, we need to assess whether observations of clusters with no halo are deep enough to detect it.
Placing upper limits on the flux from extended structures in interferometric images is not a trivial task, and different methods have been used in the past.
\citet{Venturi08}  have used the injection of mock radio halos in the UV-plane. We have followed their approach, modifying the morphology of the mock radio halos to resemble the observed ones.

\subsubsection{Mock halos: expected properties}
The power of radio halos is known to correlate with the mass of the host cluster \citep[e.g.][]{Cassano13,SommerBasu13}. More massive clusters host more powerful radio halos,
in line with the idea that the source of energy for the radio emission is the gravitational energy released in the ICM during mergers.
To predict the expected power of radio halos at 1.4 GHz (\Pradio), we have used the correlation between  \Pradio  and  $M_{500}$ by \citet{Cassano13}. We note that
this correlation has been derived using massive clusters only ($\rm M_{500}>  4 \times 10^{14} M_{\odot}$). A tighter correlation has been published by \citet{Martinez16}, while we were writing this paper.
Although slightly steeper than the one by \citet{Cassano13}, the two are consistent within their 1$\sigma$ errors.\par
Another correlation exists between  \Pradio and the size of the radio halo \citep{GiovanniniFeretti2000}.
In particular, \citet{Cassano07} have found a direct scaling between \Pradio and $R_{\rm H}=\sqrt{R_{\rm min} \times R_{\rm max}}$, where $R_{\rm min}$ and $R_{\rm max}$ are the minimum and  maximum radii of the radio halos, respectively.\par
\citet{Murgia09} have analysed a sample of radio halos and they have found that the azimuthally averaged brightness profile, $I(r)$, is well represented by an exponential law:
\begin{equation}
\label{eq:radioprofile}
I(r)=I_0e^{-r/r_e},
\end{equation}
where $r_e$ and $I_0$ are two independent fit parameters. We have compared the values of $R_{\rm H}$ and $r_e$ found by \citet{Cassano07} and  \citet{Murgia09}
for the eight clusters in common in their samples. The median value of the ratio $R_{\rm H}/r_e$ is 2.6.\par
We used the \PMcorrelation correlation by \citet{Cassano13} to predict the expected power of the radio halos at 1.4 GHz. Then, we used the correlations by 
\citet{Cassano07} and \citet{Murgia09} to predict the expected size of the radio halos ($R_{\rm H}$ and $r_e$).\par
Although the radio halo brightness can be described by azimuthally averaged profiles, the morphology of radio halos is irregular and brightness fluctuations are present on a range of spatial scales. To account for this, we have modelled the mock halos as 2D fields with power spectrum fluctuations. The power spectrum follows the form $P(\Lambda)\propto \Lambda ^{n}$, where $\Lambda$ is the spatial scale and ranges between 10- 250 kpc, and $n=11/3$. As discussed below, observations do not permit to constrain the range of 
$\Lambda$ and the value of $n$. Nonetheless, constraints about cluster magnetic fields indicate that the numbers we have adopted here are reasonable \citep{Govoni05,Govoni06,Bonafede09b}. Then, the radio profile has been normalised to follow Eq. \ref{eq:radioprofile}. This way, we obtain models of the radio halos with the radio power and size that obey known correlations and include fluctuations in the surface brightness distribution.

\subsubsection{Imaging of mock halos and determination of upper limits}
Once the parameters of the model halos are set, we have derived upper limits on the radio halo power as described below:
\begin{enumerate}
\item[1)]{To start with, $I_0$ has been chosen such that the radio power within 2.6 $r_e$ (i.e. $R_{\rm H}$) follows the  \PMcorrelation correlation. The flux density of the radio halo within $R_{\rm H}$ is referred as $\Smt$.}
\item[2)]{The images of the mock halos have been Fourier-transformed and added to the visibilities of the observations. 
The positions of the mock halos have been chosen to be close to the cluster centres and in regions devoid of radio sources.
The new datasets have been imaged and deconvolved with the software package CASA 4.5 \citep{CASA}, using the Briggs weighting scheme \citep{briggs} with different robust parameters, and UV-tapers to achieve a beam of $\sim$40-50 arcsec.}
\item [3)]{The largest detectable size within a circle of radius $R_{\rm H}$ and above 2$\sigma$ has been measured ($\Dmm$). We considered the mock halo as detected if $\Dmm \geq R_{\rm H}$, i.e. if the largest detectable size above 2$\sigma$ is at least half of the expected halo size. We define $\Smme$ as the flux density measured within $\Dmm$.
}
\item[4)] {If $\Dmm \gg R_{\rm H}$, or $\Smme > 30\% \Smt$, 
 we have  repeated the steps above decreasing the value of $I_0$ by 0.1 $\mu$Jy/arcsec$^2$, to see if we can achieve a more stringent limit.  Conversely,  if $\Dmm <  R_{\rm H}$, we have repeated steps above increasing the value of $I_0$ by 0.1 $\mu$Jy/arcsec$^2$.}
\item[5)]{steps 2-4 are repeated until the mock halo is considered detected, according to the conditions explained above.
For the upper limits that we have computed, the measured $\Smme$  is 30 - 70\% of  $\Smt$.}
\end{enumerate}
We show in Fig. \ref{fig:A2345} a series of images of mock halos with increasing brightness. \par
The upper limits for the clusters A3376 and MACSJ0025 have been derived using data at 325 MHz, assuming that the size of the radio halo would not change significantly between 325 MHz and 1.4 GHz, and assuming $\alpha=1.3$.\par
As our main criterium is based on $\Dmm \geq R_{\rm H}$, this approach is very conservative for clusters with $M_{500} \geq 7.5 \times 10^{14} M_{\odot}$. For these objects, the expected $R_{\rm H}$ is $>$ 500 kpc, while in reality a source larger than 500 kpc would already be classified as a radio halo. This is not critical for the present study, as there is only one cluster in our sample with $M_{500} > 7.5 \times 10^{14} M_{\odot}$, but it will become important when larger samples become available.\\

 Similarly, the upper limits on \Pradio for clusters with $M_{500} \leq 3.4 \times 10^{14} M_{\odot}$ should be treated with caution. For these clusters, the estimated size of the radio halos would be $< 500 $ kpc, and the source would likely not be classified as a radio halo. As a consequence, the limits on the power of the radio halos in A3365, A3376, and ZwCl0008 should be treated carefully. For our analysis, this is not crucial as the limits we have assumed are consistent with the \PMcorrelation correlation.\par

\subsection{Upper limits and the \PMcorrelation correlation}
Clusters with and without radio halos occupy separate regions of the \PMcorrelation correlation \citep{Cassano13}. As mentioned before, 
most of the clusters with radio halos are in merging systems, while most of the upper limits are found for non-merging clusters.
However, not all massive merging clusters host a radio halo, and this raises questions about their origin. \par
\begin{figure}
\includegraphics[width=\columnwidth]{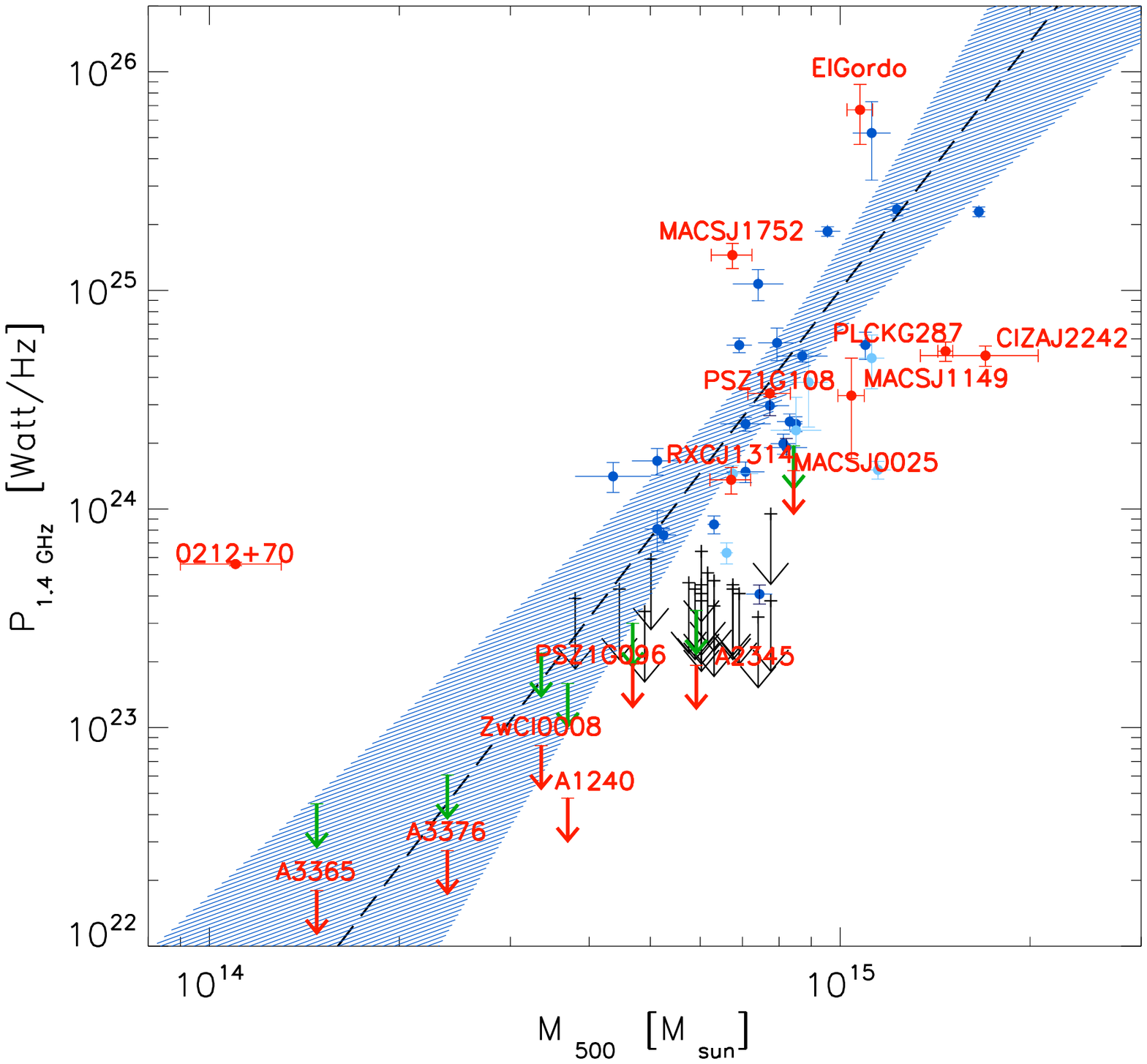}
 \caption{Radio power at 1.4 GHz versus cluster mass within $\rm R_{500}$. Filled dots are radio halos: light blue dots mark halos with a spectrum steeper than $\alpha=1.5$ \citep[][]{Venturi07,Venturi08,Cassano13,Bonafede14b,Bonafede15}, red dots mark halos in clusters with double relics. Black arrows are upper limits on $\rm P_{1.4 \, GHz}$ \citep[][and ref. therein]{Venturi08,Cassano13}. The dashed line is  the \PMcorrelation correlation as derived by \citet{Cassano13} excluding steep-spectrum halos.  The shaded area marks the 95\% CL of the correlation.
 Red and green arrows are upper limits on halos in clusters with double relics derived in this work, computed from $\Smt$, and $\Smme$, respectively (see text for more details). Clusters analysed in this work are labelled.  }
 \label{fig:UL}
\end{figure}

In Fig. \ref{fig:UL}, the  \PMcorrelation correlation is shown, together with the upper limits derived here. Both, the upper limits corresponding to $\Smme$and  $\Smt$ are plotted in Fig. \ref{fig:UL}.  If we assume that the power of the radio halos on the \PMcorrelation correlation represents the whole radio emission in the cluster, i.e. that all the radio emission has been detected, $\Smt$ are the values that we have to consider. It is more likely that a fraction of the halo flux has been missed by observations, and a more realistic value for the upper limit should lie in between $\Smt$ and $\Smm$. Here, we go here for a conservative approach and consider  $\Smt$.\par
From the analysis in the previous section, we conclude that three or four clusters with double relics and no previously detected radio halo do not host a radio halo that follows the \PMcorrelation correlation. They are PSZ1 G096.89+24.1, Abell 2345, and MACSJ0025.4-1222. For the cluster A1240, we can put an upper limit below the correlation only if we consider the
flux density that is measured in the image ($\Smm$), while the corresponding flux density in the model halo ($\Smt$) would still be consistent with the correlation within 95\% confidence level (the shaded area in Fig. \ref{fig:UL}).
For Abell 3376, ZwCl0008, and A3365 we can only place upper limits consistent with the \PMcorrelation correlation within 95\% confidence level.\par

As  \Pradio $\propto M_{500}^{3.77}$ while  \Pradio $\propto R_{H}^{4.18}$ \citep{Cassano07, Cassano13}, the radio brightness is proportional to $M^{1.97}_{500}$. This explains why we are able to place lower upper limits for clusters with smaller $M_{500}$ even if observations are not deeper (see values of $I_0$ listed in Table \ref{tab:NH}). \par

Note that the limits that we have derived depend on the model that we have assumed for the radio halos.
The choice of some parameters for the mock halo modelling is arbitrary, 
as present observations do not allow us to study the spectrum of the halo brightness fluctuations. Nonetheless, the approach that we used has two improvements with respect to previous methods: (i)
The fluctuations that we introduce in the halo brightness distribution make the mock halo appear more like the observed ones.
(ii) The radial profiles follow the azimuthally averaged brightness profiles that have been observed.

\begin{figure*}
\includegraphics[width=17cm]{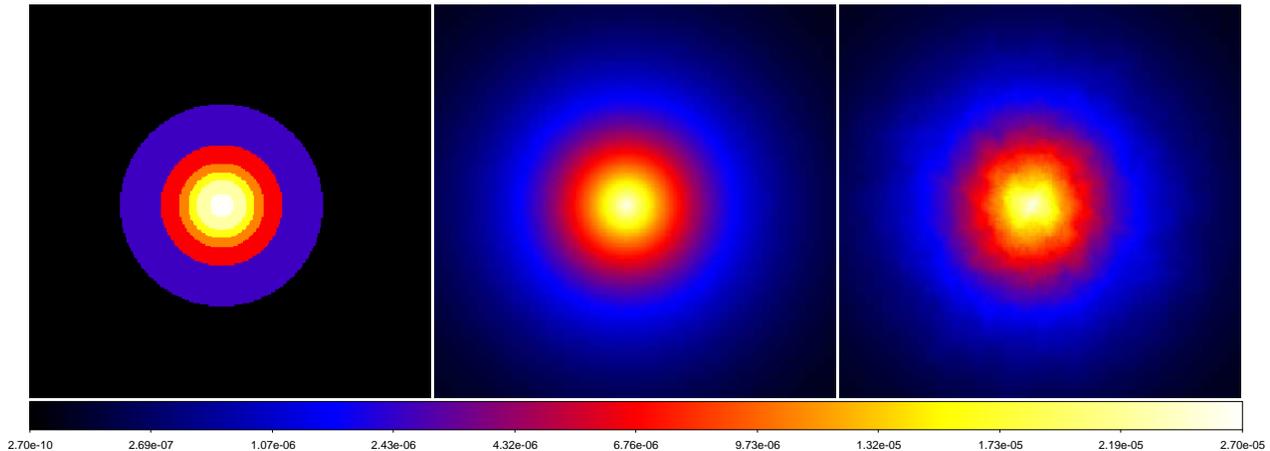}
 \caption{Models for the mock halos used to derive upper limits. Left: models with concentric spheres \citep{Venturi08}, middle: model with exponential profile, left: model with average exponential profile and power spectrum fluctuations. The three models have the same flux density within the same radius.  }
 \label{fig:models}
\end{figure*}

\subsection{Comparison with the previous method}
The idea of injecting mock radio halos in the UV-plane to derive limits on the radio halo power has been introduced by \citet{Venturi08}.
Here, we compare our method to the one used by \citet{Venturi08} to assess the impact on the upper limits.
For this comparison, we take two clusters:  Abell 2345 and Abell 3376. 
\begin{itemize}
\item A2345 is chosen because the images have the most uniform noise pattern among the
clusters analysed here. Following \citet{Venturi08}, we have modelled mock halos as concentric circles of increasing radius. The largest circle
has a radius $=R_{\rm H}$, and contains $\sim 50\%$ of the halo flux density. 
We have injected a radio halo with the same power at 1.4 GHz as the one listed in Tab. \ref{tab:NH}, corresponding to the upper limit we have derived for A2345.
We found that the flux recovered in the two cases (concentric circles and exponential profile) measured above 2$\sigma$ is the same within the errors.
Moreover, the largest linear size of the mock halo, measured above 2$\sigma$, is $\sim$6\% smaller when concentric spheres are used. Hence, we can conclude that in this case differences are not significant. \par
This might indicate that the dominant factor is not the mock halo brightness distribution, but a combination of
i) the number of independent beams with which the halo is sampled, which depends on the FWHM of the restoring beam and on the angular size of the mock radio halo;
ii) the halo surface brightness, which depends on the observing frequency.\par
As halos have a steep spectrum, their surface brightness decreases as the observing frequency increases. At 1.4 GHz, they are better imaged with large beams, which could smooth out
the differences of the halo models.  In the case of A2345, the data are taken at 1.4 GHz, the mock halo has a brightness of 0.6 mJy/beam and it is sampled  by 18 beams (4 in the area above $2\sigma$).\par
\item We have repeated the same comparison for the cluster Abell 3376, that has been observed at 325 MHz.
The halo is sampled by 26 beams (8 in the area above $2\sigma$), and it has a mean brightness of 4.6 mJy/beam, which is the highest among the clusters analysed here. As expected, the differences in this case are larger: the halo modelled with concentric spheres enables recovery of a higher flux density above $2\sigma$. Specifically, we recover 12\% more flux above 2$\sigma$, and the halo angular size measured above 2$\sigma$ is $\sim$7\% smaller.
This can be understood as the model with concentric spheres has a profile that is more peaked towards the centre with respect to the models with power-spectrum fluctuations and exponential profile (see Fig. \ref{fig:models}).
\end{itemize}
Our analysis indicates that as the halo brightness increases (e.g. the halo is observed at lower frequencies), the upper limit to the
radio power at 1.4 GHz depends on the modelling of the mock halo. \par
Analysing  how the limits on \Pradio are affected by observing frequency, halo angular size, models of the mock halo, and many other parameters, goes beyond the aim of this work. 
 However, instruments like LOFAR and the future SKA precursors in the coming years, are able to image extended emission at high angular resolution and sensitivity. Hence, we expect that upper limits derived from observations of those instruments will depend on the modelling of the mock halos, at which time a more thoughtful investigation of these parameters should be done.
 
\subsection{Upper limit and halo model assumptions}

\subsubsection{Power spectrum brightness fluctuations}
We investigate in this section the impact of models with and without power-spectrum fluctuations on the upper limits.
As above, we have used the clusters A2345 and A3376. 
For A2345, we found that the flux density measured in the two cases is consistent within the errors, although higher when power spectrum fluctuations are not included. 
 In the case of A3376, the flux measured is 6\% higher if power spectrum fluctuations are not included. Although negligible, these differences demonstrate that future observations at low frequency and high resolution will likely be sensitive to the modelling of the mock halo.

\subsubsection{Ellipsoidal halos}
Some clusters with double relics have a radio halo that extends in-between the two relics. This is the case of CIZAJ2242, MACS1752, and El Gordo. For these clusters, the radio halo would be better represented by an ellipsoid with the major axis equal to the distance between the two relics. \par
We have compared the value of $R_{\rm H}$ measured for the halos in CIZAJ2242, MACS1752, and El Gordo with the one predicted by \citet{Cassano07}. For the clusters CIZAJ2242 and ELGordo, the values are consistent, while for MACSJ1752 the measured value of $R_{\rm H}$ is almost twice the predicted one. \par
In order to estimate how the upper limits change if the mock halos are modelled as ellipsoids, we consider two scenarios in which the radio halo does not follow the $M_{500}-R_{\rm H}$ correlation and the exponential profile in Eq. \ref{eq:radioprofile}. 
\begin{enumerate}
\item{The mock halo surface brightness profile follows an exponential law of the form $I=I_0 e^{- \epsilon / r_e}$, with $ \epsilon=\frac{x^2}{a^2}+\frac{y^2}{b^2}$ for $\epsilon \leq 1$}
\item{The mock radio halo has a constant surface brightness.}
\end{enumerate}
In both scenarios, the halo is modelled as an ellipsoid with the major axis equal to the distance between the two relics and the minor axis equal to the relics size. We have assumed that the \PMcorrelation  correlation holds, i.e. the total flux of the model halo is the same as the mode in Sec. \ref{sec:UL}.\par
Scenarios (i) and (ii) can be considered as the most extreme cases of halos that follow the exponential fit, although not spherical but elliptical (i), and halos that do not show any brighter part.
We have derived the upper limits for the cluster Abell 2345 in both scenarios. It results that for model (i) the upper limit is consistent with the one we put in Sec. \ref{sec:UL}, while for model (ii) the upper limit would be consistent with the \PMcorrelation  correlation. \par
The analysis that follows in the next section will assume that halos can be modelled as done in Sec. \ref{sec:UL}.

\begin{table*}
 \centering

 \caption{Clusters with no halo}
 \begin{tabular}{lcccccccc|}
 
  \hline
Cluster			&   Expected \Pradio &   \Pradio injected &  \Pradio meas & $r_e$  & $I_0$     & $\Dmm$ & $2 \times R_{\rm H}$\\
					&   W/Hz               	&   W/Hz                                &    W/Hz                     & kpc     &  $\mu$Jy/arcsec$^2$ &  kpc & kpc  \\
&&&\\
Abell 3365			& $6.4 \times 10^{21} $ 	& $4.5 \times 10^{22} $ & $1.8 \times 10^{22} $ & 45 &  1.5  & 170 & 230  \\
Abell 3376	$^{*}$	& $3.8 \times 10^{22} $ 	& $6.8 \times 10^{22} $  & $2.7 \times 10^{22} $&  69 & 3.0  &  210  & 360\\
ZwCl0008.8+5215 		& $1.4 \times 10^{23} $ 	& $2.4 \times 10^{23} $ &	 $8.3 \times 10^{22} $ &94   & 0.7  & 365  & 490\\
Abell1240				& $2.0 \times 10^{23} $ 	& $1.6 \times 10^{23} $  & $4.8 \times 10^{22} $  & 103 & 0.3  & 300  & 530\\
PSZ1 G096.89+24.1		& $4.9 \times 10^{23} $ 	& $3.8 \times 10^{23} $ & $2.0 \times 10^{23} $ &127     & 0.3  & 400 & 660\\
Abell 2345			& $1.2 \times 10^{24} $ 	&  $3.8 \times 10^{23} $ &  $1.9 \times 10^{23} $  & 157& 0.3  & 480  & 810 \\
MACSJ0025.4-1222$^{*}$	& $4.5 \times 10^{24} $ 	&$ 1.9\times 10^{24} $  & $ 1.5\times 10^{24} $&  216  & 1.5   &  800  & 1130\\

\hline
\multicolumn{8}{l}{\scriptsize Col. 1: cluster name; Col. 2: expected \Pradio according to the \PMcorrelation  correlation \citep{Cassano13}; Col. 3: \Pradio of the model   }\\
\multicolumn{8}{l}{\scriptsize injected in the UV-data (corresponding to $S_{R_{\rm H}}^{mock, tot}$); Col. 4: \Pradio measured in the image (corresponding to $\Smme$); Col. 5, 6: parameters }\\
\multicolumn{8}{l}{\scriptsize of the mock halo corresponding to the upper limits. Col. 7: Size of the mock radio halo measured in the image above $2\sigma$; Col. 8: expected size }\\
\multicolumn{8}{l}{\scriptsize of the radio halo according to the \Pradio - $R_{\rm H}$ correlation \citep{Cassano07}.}\\
\multicolumn{8}{l}{\scriptsize *=upper limit derived at 325 MHz, \Pradio  have been computed assuming $\alpha=1.3$}\\
\end{tabular}
\label{tab:NH}
\end{table*}

\begin{figure}
\includegraphics[width=0.9\columnwidth]{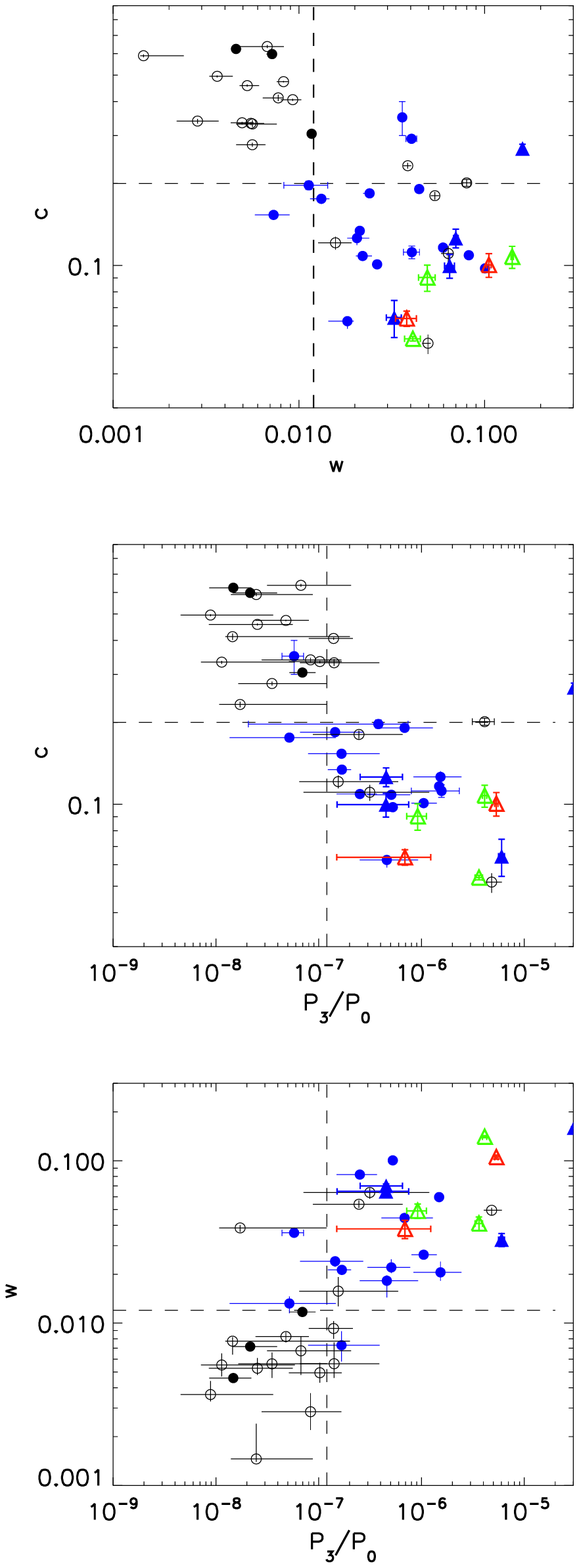}
 \caption{Morphological estimators of cluster dynamical status from X-rays. Black filled dots are clusters with mini halos, blue filled dots are 
 clusters with radio halos and no double relics (from \citealt{Cassano10,Venturi08,Bonafede14b,Bonafede15}), blue filled triangles are clusters with radio halos and double relics, red open triangles are clusters with double relics and no radio halo (upper limit below the \PMcorrelation correlation), black spheres are clusters with no radio halos and no double relics. Green triangle mark clusters where no radio halo has been detected but the upper limit is above the \PMcorrelation correlation. Dashed lines are the median of the parameters, that define radio-loud and radio-quiet quadrants \citep{Cassano10}}
 \label{fig:xray}
\end{figure}

\section{Merging status from X-ray data}
\label{sec:X}

In this section, we analyse the X-ray properties of the clusters with double relics and compare them to other clusters --- with and without radio halos --- analysed in the literature. In our sample, there are nine clusters with {\it Chandra} archival observations, that can be suitable for a comparison with literature information. These clusters are listed in Table \ref{tab:sample}. 

\subsection{X-ray morphological indicators}
X-ray data can be used to assess the dynamical status of a cluster through three morphological indicators \citep[e.g.][]{Boehringer10}:
\begin{itemize}
\item{The concentration parameter c, defined as 
\begin{equation}
 c= S_x ( r < 100 {\rm kpc}) / S_x (r< 500  {\rm kpc}),
 \end{equation}
 
 with $S_x$ being the  integrated X-ray surface brightness and $r$ the radial coordinate from the cluster center.}\par
\item{The power ratio $P_3/P_0$. The power ratio is a  multipole decomposition of the projected mass distribution inside a given radius (which we define as $R_{ap}=$500 kpc). 
As the gas is subject to the cluster gravitational potential, the idea behind the power ratio analysis is that the X-ray surface brightness can be used to trace the projected mass distribution.

\cite{Boehringer10}, found that $P_3/P_0$ is the lowest power ratio moment that can give a measure of the cluster substructures. $P_0$ is defined as 
\begin{equation}
P_0 = S_x(r<R_{ap}) \ln(R_{ap}),
\end{equation}
while 
\begin{equation}
P_3=\frac{1}{18  R_{ap}^6} (a^2+b^2),
\end{equation}
with 
\begin{equation}
a=\int _{r<R_{ap} }{S_X(x') r'^3\rm{cos}(3\phi') d^2x'},
\end{equation}
\begin{equation}
b=\int_{r<R_{ap}}{S_X(x') r'^3 \rm{sin}(3\phi') d^2x'}.
\end{equation}
The higher the value of $P_3/P_0$, the more substructures are present in the X-ray surface brightness. Hence, high values of $P_3/P_0$ indicate a merger cluster;}
\item{The centroid-shift parameter, $w$, measures the standard deviation of the projected separation between the X-ray peak and the centroid in units of $R_{ap}$, computed in $N$ spheres of increasing radius. Specifically, it is defined as:
\begin{equation}
w=\sqrt{  \frac{1}{N-1}  \Sigma (\Delta_i - \langle \Delta \rangle)^2  } \times \frac{1}{R_{ap}},
\end{equation}
where $\Delta_i$ is the distance between the cluster centre and the centroid of the $i-$th circle. We start from $R=0.05 R_{ap}$ and increase the radius at steps of $0.05 \times R_{ap}$ until 
$R=R_{ap}$. As for  $P_3/P_0$, high values of $w$ indicate that the cluster is far from dynamical equilibrium.
Using numerical simulations, \citet{Poole06} has found that $w$ is very sensitive to the cluster dynamical state. }
\end{itemize}
\citet{Cassano10} have computed these indicators for a mass-selected sample of clusters, and they found that clusters with and without radio halos occupy different regions
in the morphological diagrams. Although some outliers have been found later \citep{Bonafede14b,Bonafede15,Sommer17,Venturi17},  most of the halos are in clusters with low $c$, high $w$ and high $P_3/P_0$.

\subsection{X-ray data reduction and analysis}
{\it Chandra} observations have been reduced using
the Chandra Interactive Analysis of Observations (CIAO) 4.7. 
The event files have been reprocessed to apply the latest calibration files as of February 2016 (v 4.7.2). 
Observations have been processed in the energy band $0.5-7$ keV, to search for soft proton flares. Time periods affected by proton flares have been excluded (usually less than 10\% of the observing time) and clean event files have been created. The clean event files have been exposure corrected and images in the band $0.5 - 2.4 $ keV have been created. Exposure-corrected images have been binned to achieve a common resolution of
4 kpc/pixel. This is necessary because we want to compare clusters at different redshifts with the sample analysed in \citet{Cassano13} at $z=0.2-0.3$, and a different resolution may impact the value of $P_3/P_0$ and w.
Point sources have been detected using the {\it wavedetect} script in CIAO. This script correlates the images with wavelets of different scales, and searches the results for significant correlations. Results from {\it wavedetect} have been inspected visually and corrected, when needed. Point sources have been excluded from the analysis.\par
The morphological parameters $P_3/P_0$, $w$ and $c$ have been computed as described above. 
In Fig. \ref{fig:xray}, we show the positions of the clusters in the $P_3/P_0-c$, $w-c$, and $P_3/P_0 - w$ diagrams (taken from \citealt{Bonafede15}, adapted from \citealt{Cassano13}). Clusters with double relics are added with different symbols: red empty traiangles
are clusters with double relics and no halo (MACSJ0025 and A2345), green empty triangles are clusters with double relics and no halo, but with upper limits on \Pradio consistent with the \PMcorrelation correlation (A1240, A3376, and ZWCl0008), blue triangles are clusters with double relics and halo (El Gordo, CIZA2242, J0212, and MACSJ1149) .
 As expected --- given the presence of double relics --- the clusters we analyse in this work are all in the merging quadrant of the plots. In addition, we note that they are characterised by $P_3/P_0$ and $w$ values among the highest of the cluster sample. This indicates that the amount of substructures and asymmetries in the mass distribution are both very high. We also note that double relic clusters with and without radio halo occupy the same region of the plots, indicating that the merging status is similar. Projection effects could play a role, as double relics should be better observable when the merger takes place in the plane of the sky, and this is the most favourable configuration to detect as many substructures and asymmetries in the gas distribution as possible. Nonetheless, it is interesting to note that clusters with double relics lie in the most disturbed region of the plots.\par

According to the X-ray substructure analysis, it remains unexplained why some of the clusters with double relics do not have  a radio halo. \par

\begin{figure}
\includegraphics[width=0.8\columnwidth]{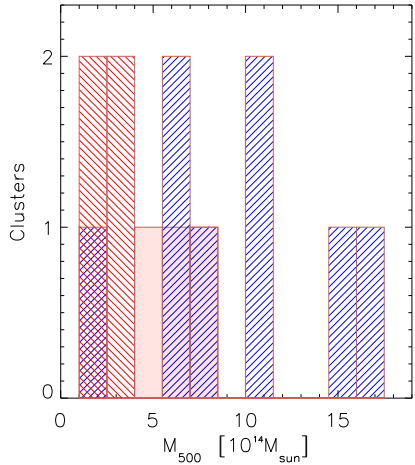}
 \caption{Histogram of the clusters with halo (blue diagonal lines), without halo (filled red), and with upper limit consistent with the \PMcorrelation correlation (red  diagonal lines) versus the cluster mass. }
 \label{fig:histo}
\end{figure}

\begin{figure}
\includegraphics[width=0.9\columnwidth]{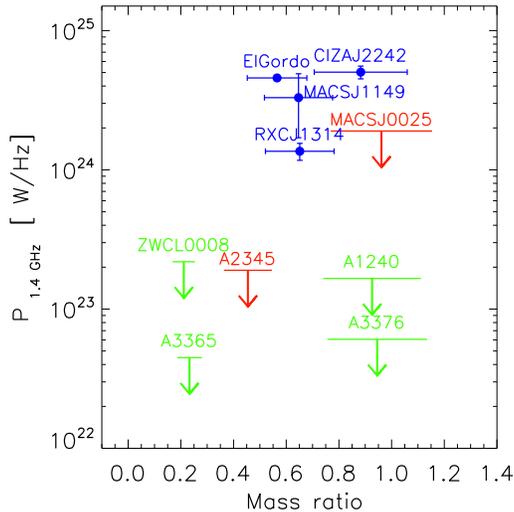}
\caption{\Pradio versus the merger mass-ratio.  Blue points are detected halos, red arrows are upper limits below the \PMcorrelation correlation,
green arrows are upper limits above the \PMcorrelation correlation. As errors on the masses of the sub-clusters are not available for most objects, we assume a fiducial error of 20\% on the mass ratios.}
\label{fig:massratio}
\end{figure}

\begin{figure}
\includegraphics[width=\columnwidth]{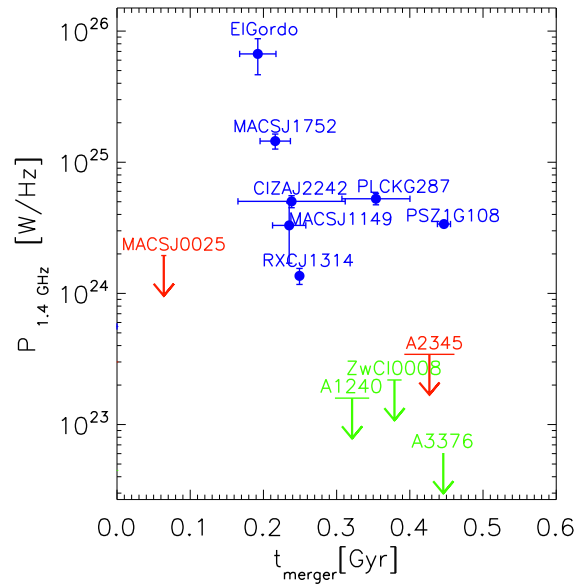}
\caption{\Pradio versus estimate of the time-since-merger. Blue points are detected halos, red arrows are upper limits below the \PMcorrelation correlation,
green arrows are upper limits above the \PMcorrelation correlation.}
\label{fig:timescale}
\end{figure}

\section{Discussion}
\label{sec:Discussion}
Our analysis shows that all clusters with double relics are merging clusters with a
similar dynamics and  that is also similar to clusters with radio halos.
However, some of them do not have a radio halo that follows the \PMcorrelation correlation. 
Although turbulent re-acceleration models make statistical predictions that are consistent with data so far, individual cases remain to be understood. This indicates that our understanding of the process is incomplete.\par
It has been suggested that the merger mass-ratio and/or a time-scale argument do not play a role \citep[e.g.][and ref. therein]{Cassano16}.
In the following analysis, we will use the Wilcoxon-Mann-Whitney  statistical test to check whether clusters with and without radio halos can be regarded as different 
groups. The Wilcoxon-Mann-Whitney test is a non-parametric test that 
 checks whether two samples are consistent with the same population or not. Specifically, the test can check --- given two independent samples --- whether one variable tends to have values higher than the other.
 It is a good alternative to the t-test when data are not normally distributed, and can be used when the sample sizes is small. 
 For more details, we refer to \citet{WMW_test}.
As independent samples, we have taken  the masses of clusters with and without radio halos, and tested the null hypothesis that the two population are not statistically different.

\subsection{Total cluster mass}
In Fig. \ref{fig:histo}, we show a histogram of the halo/no halo distribution versus the cluster mass. 
This plot hints at a mass trend, as clusters without radio halo tend to lie in the small mass region.
For a quantitative test, we have run the Wilcoxon-Mann-Whitney  statistical test. To start with, we have run the test including in the ``no-halo" sample only the clusters for which we have put upper limits below the correlation. The null hypothesis that the 
the clusters are drawn from the same population cannot be rejected at any of the confidence level tabulated. We have repeated the test including all the clusters with upper limits in the ``no-halo" sample, and found again that the null hypothesis cannot be rejected. 

\subsection{Mass ratio}
Clusters with double relics and with or without radio halos could trace mergers with different mass ratios. Recently, \citet{Cassano16}  have analysed a sample of massive clusters and derived that the fraction of merger clusters is higher than the fraction of clusters with radio halos. This could suggest that radio halos trace mergers with larger mass ratios. 
From the literature, we have collected the data about the masses of the clusters in our sample. Estimates come from strong and weak lensing analysis 
(ZwCl0008: \citealt{Golovich17}, CIZAJ2242: \citealt{Jee15}, ElGordo: \citealt{Jee14}, MACSJ0025: \citealt{Bradac08}), 
from dynamical analysis (MACSJ1149: \citealt{Golovich16}), and from cluster members velocity dispersion (A2345 \citealt{Barrena11},  RXCJ1314, A1240 , A3376 , A3365: Golovich et al. in prep.). We have defined the mass ratio as $M_{\rm min}/M_{\rm max}$, where $M_{\rm min}$ and $M_{\rm max}$ are the minimum and maximum masses of the two main sub-clusters. In Fig. \ref{fig:massratio}, \Pradio is plotted versus the mass ratio. Although the sample is small, data do not seem to support a different mass-ratio for clusters with and without radio halos.
To quantify this, we have run the  Wilcoxon-Mann-Whitney statistical test. As independent samples, we have taken  the mass ratios 
derived for clusters with and without radio halos. We have repeated the test twice, first considering in the sample of clusters without radio halos only the upper limits below the \PMcorrelation correlation (red arrows in Fig. \ref{fig:massratio}) and then including also the upper limits that would be consistent with the \PMcorrelation correlation (green arrows in Fig. \ref{fig:massratio}).
In both cases,  the null hypothesis that the clusters are drawn from the same population cannot be rejected. Hence, no indication is given by present data about a different radio behaviour for cluster mergers with different mass ratios.\par

\subsection{Time since merger}
If halos are powered by turbulence injected in the ICM during mergers, we may expect that the cascade takes time to develop and enter in the regime where it can accelerate particles.
Hence, clusters with double relics and no radio halos could be those undergoing a merger more recently than clusters with double relics and halos.\par
Alternatively, clusters with relics and no halo on the \Pradio-$\rm M_{500}$ relation could be those that have undergone a merger a longer time ago. While shock acceleration takes place at relics, the halo emission at 1.4 GHz is fading quickly due to synchrotron and inverse Compton losses. In both cases (earlier or later merger), the two classes of clusters should be divided by a characteristic time since the merger took place.
 Obtaining an estimate of the time since merger started is difficult and little data is available in the literature. 
However, we can take the distance between the two relics ($d_{r,max}$) as a proxy of the time since the merger happened. To do this, we made the following assumptions:\\
$i)$ The ICM is isothermal;\\
$ii)$ Relics trace two symmetric shock waves injected in the ICM at a given time;\\
$iii)$ Radio relics are powered by Diffusive Shock Acceleration (DSA);\\
$iv)$ The Mach number of the shock wave has a shallow radial dependence $M \propto r^{1/2}$ \citep[][and ref. therein]{va09shocks,Vazza10,Hong14}  ;\\
$v)$ The merger is taking place in the plane of the sky (i.e. projection effects are minimal), and both shocks travelled the same distance equal to half of the distance of the two relics;\\
We have estimated the cluster temperature, $T$, and the sound speed, $c_s$, in the ICM using the  $M_{500}-T$ correlation by \citet{Pratt09}.
Assuming iii), we can derive the Mach number  ($M_{\rm shock}$) of the shock wave from the radio spectral index ($\alpha$, \citealt{Drury83}). From $M_{\rm shock}$ and $c_s$, we derived the shock velocity $v_{\rm shock}$ as the shock wave propagated outwards, and the time that the shock wave spent to arrive at the position of the relic ($t_{\rm merger}$).
Values of $d_{r,max}$  and $\alpha$ have been taken from \citet{deGasperin14,deGasperin15,Riseley17}. We refer to Table \ref{tab:sample} for references on single objects. Unfortunately, no information of $\alpha$ is available for the clusters A3365, 0212+70, and PSZ1 G096.89+24.1. 
 When $\alpha \leq 1$, we have assumed that the spectral index is the injection spectral index. Otherwise, we have assumed that the injection spectral index can be derived by flattening the integrated spectral index by 0.5.\par

We have computed $t_{\rm merger}$ independently for each relic. In Fig. \ref{fig:timescale}, we plot the mean of $t_{\rm merger}$ for each cluster versus radio halo power. If the time since the merger started is the key quantity to switch on the radio emission in merging clusters, clusters with and without radio halos should lie in different regions of the plot. 
We note that A2345 and MACSJ0025 (red arrows in Fig. \ref{fig:timescale}) lie at early and late merger with respect to clusters with double relics and a radio halo. Also clusters with double relics for which we put an upper limit to \Pradio consistent with the correlation (green arrows in Fig. \ref{fig:timescale}) seem associated to late mergers. 
However, the sample is too small to draw any conclusions, and no suitable statistical test can help us assessing whether $t_{\rm merger}$ plays a role in the formation of the radio halo\footnote{The Wilcoxon-Mann-Whitney test cannot be used in this case, because we do not want to test whether one variable tends to have values higher than the other, but we want to test whether one variable has a different distribution. Other test exist in this case, but to our knowledge none of them can be used for small samples.}.\par
It must be noted that the timescale  $t_{\rm merger}$ that we derive here is not the time after core passage ($t_{\rm cross}$) that is derived through more accurate modelling (see e.g. \citealt{Golovich17}).
Unfortunately, data do not allow us to derive estimates of $t_{\rm cross}$ for all the clusters in the sample. The $t_{\rm merger}$ estimates that we have derived are the best estimate that current data can provide, but could be affected by the assumptions i) - iv).
Once more data are available, the analysis should be repeated using estimates of $t_{\rm cross}$ from lensing analysis, to better evaluate the role of the time since core passage.\par
\smallskip
It must be noted that we are dealing with very small samples. Although non-parametric statistical tests can deal with samples as small as those we have here, our hope is that future radio surveys will increase the sample of clusters with double relics and enable lower upper limits in order to investigate the role of timescale, mass, and mass ratio.

 \section{Conclusions}
 \label{sec:Conclusions}
In this work, we have analysed all clusters with double relics known to date, in order to understand why radio halo emission is found in only a fraction of them. Our results can be summarised as follows:
\begin{itemize}
\item{We have developed a new procedure to derive upper limits on the halo emission.
 We have modelled the radio halos as exponential functions, as observed by \citet{Murgia09}, and with sizes that follow the observed correlation
 between \Pradio and $R_H$ \citep{Cassano07}. We also added brightness fluctuations in a range of spatial scales (10 - 250 kpc) to resemble observed radio halos. With our method, we have placed upper limits below the \PMcorrelation correlation for three or four clusters.}
\item{We have recomputed the upper limits to \Pradio  assuming that clusters with double relics do not follow the observed properties of radio halos, to test the robustness of our limit. In particular, we have  modelled the halos as ellipsoids with sizes much larger than expected from the \Pradio - halo size correlation. We found that if halos have a constant brightness, no upper limit can be put below the \PMcorrelation correlation. Instead, our limits still hold if the brightness distribution is exponential.  }
\item{Our analysis indicates that the large beams needed to recover the diffuse emission in data published so far are likely to smooth out the differences in the halo models.
However, differences between this new method and previous ones become significant if clusters are observed at low frequency, high resolution and sensitivity. 
Hence, this method is promising to derive deep upper limits with the new generation of radio instruments, such as LOFAR and the SKA pathfinders.}
\item{Using {\it Chandra} archival data, we have computed the cluster morphological parameters. As expected, clusters are placed in the {\it merger region} of the morphological diagrams, irrespective of the presence or absence of radio halos. Although projection effects could play a role, clusters with double relics show the highest levels of disturbances in the X-rays. Hence, we conclude that based on the X-ray analysis it remains unexplained why only some clusters with double relics have a radio halo.}
\item{Using literature information and under simple assumptions on the clusters properties, we have investigated whether the the presence/absence of a radio halo depends on different timescales since merger happened  ($t_{\rm merger}$) or on different mass ratios of the merger sub-clusters. 
Althought the sample of clusters is small, data do not suggest a dependence on the mass ratio: the null hypothesis that clusters with and without radio halos can be interpreted as different samples depending on the merger mass-ratio cannot be rejected.
We note that clusters without halos tend to be associated with late and early mergers, but data are too scarce to perform a statistical test.}
\end{itemize}
A different magnetic field in clusters with and without radio halos could also explain the presence/absence of radio emission. Data are scarce, but do not support this scenario \citep{Bonafede11a}. In addition, it would be hard to understand from a theoretical point of view why clusters with similar masses and merging status should have different magnetic field properties.\par
According to re-acceleration models, clusters with $M _{500 }\sim 5-7\times 10^{14} M_{\odot} $ could host radio halos that become visible
only at low radio frequencies (around 150 MHz, see \citealt{Cassano13}). 
These radio halos should be produced by less-energetic mergers, that cannot inject enough energy to accelerate particles up to $\sim$ GeV.
The upper limits that we have derived are computed from observations at 300 MHz and 1.4 GHz, and these clusters could  host a radio halo at lower frequencies. 
On going LOFAR and MWA surveys --- together with follow-up X-ray and lensing studies --- will shed light on this issue, as they are expected to detect more clusters with double relics, increasing the poor statistics we have now, and possibly revealing that either the mass-ratio and/or the time-since-merger play a role in understanding the absence of radio halos in merging clusters.

\section*{Acknowledgements}
We thank F. Vazza, D. Dallacasa, and G. Brunetti for useful discussions. RK acknowledges the support from the DST-INSPIRE Faculty Award by the Department of Science and Technology, Government of India.
 This research had made use of the NASA/IPAC Extragalactic Data Base 
(NED) which is operated by the JPL, California institute of technology under contract with the National Aeronautics and 
Space administration.

\bibliographystyle{mn2e}
\bibliography{master}

\label{lastpage}
\end{document}